# An Explainable Framework for Machine learning-Based Reactive Power Optimization of Distribution Network

Wenlong Liao[1], Benjamin Schäfer[2], Dalin Qin[3], Gonghao Zhang[3], Zhixian Wang[3], Zhe Yang[4]
[1]Wind Engineering and Renewable Energy Laboratory, École Polytechnique Fédérale de Lausanne(EPFL), Lausanne, Switzerland
[2]Institute for Automation and Applied Informatics, Karlsruhe Institute of Technology, Karlsruhe, Germany
[3]Department of Electrical and Electronic Engineering, The University of Hong Kong, Hong Kong, China
[4] Department of Electrical Engineering, The Hong Kong Polytechnic University, HongKong, China

***Abstract*—To reduce the heavy computational burden of reactive power optimization of distribution networks, machine learning models are receiving increasing attention. However, most machine learning models (e.g., neural networks) are usually considered as black boxes, making it challenging for power system operators to identify and comprehend potential biases or errors in the decision-making process of machine learning models. To address this issue, an explainable machine-learning framework is proposed to optimize the reactive power in distribution networks. Firstly, a Shapley additive explanation framework is presented to measure the contribution of each input feature to the solution of reactive power optimizations generated from machine learning models. Secondly, a model-agnostic approximation method is developed to estimate Shapley values, so as to avoid the heavy computational burden associated with direct calculations of Shapley values. The simulation results show that the proposed explainable framework can accurately explain the solution of the machine learning model-based reactive power optimization by using visual analytics, from both global and instance perspectives. Moreover, the proposed explainable framework is model-agnostic, and thus applicable to various models (e.g., neural networks).**

***Index Terms*—Distribution network, reactive power optimization, explainable artificial intelligence, data driven, Shapley value**

## I. INTRODUCTION

Reactive power optimization (RPO) plays a critical role in reducing power losses and maintaining voltage in distribution networks by adjusting the operational parameters of various power devices subjected to non-linear constraints. The significance of RPO for distribution networks lies in its integral role in the planning and scheduling process, which holds immense value for both theoretical exploration and practical implementations [1], [2].

The computational burden associated with the nonlinearity and non-convexity of the RPO model is a significant challenge. Traditional methods like linear programming and dynamic programming can partially alleviate this complexity by simplifying the model [3], but they may result in locally optimal solutions. These traditional research efforts have attempted to reduce the computational burden associated with the RPO model by compromising accuracy [4]-[6]. In addition to traditional linear or dynamic programming methods, various heuristic algorithms, such as genetic algorithm (GA) and wolf pack algorithm have been widely adopted to solve the RPO model through numerous iterations [7], [8]. Although heuristic algorithms demonstrate effectiveness in achieving robustness and high accuracy for solving the RPO model with nonconvex functions, they are also accompanied by a significant computational burden, especially for large-scale distribution networks.

Recently, a large number of machine learning models have been developed to quickly solve RPO models. Most, if not all, machine learning models fall into two groups: similarity-based methods and regression-based methods. In particular, the widely adopted similarity-based methods include random matrix theory, case-based reasoning, similarity matching, and Apriori algorithm [9], which aim to leverage historical data to gain insights, and make informed decisions based on the similarity measures between different cases. Generally, these similarity-based methods have limited accuracy, since they directly assign solutions of historical cases to new cases, which is not the best solution when there are significant differences between current load conditions and historical ones [10]. It is critical to consider the unique characteristics and contextual factors of new cases to ensure accurate and reliable decision-making. In contrast, regression-based methods can be treated as complex functions, which directly map load conditions to the RPO solutions (i.e., operational parameters of various power devices). For example, the work in [11] employs a multi-layer perceptron (MLP) with several dense layers to represent the non-linear relationship, while the research in [12] constructs a convolutional neural network (CNN) with multiple convolutional and pooling layers to extract latent features from the load conditions. Further, a variant of CNN called capsule network is presented in [13] to capture temporal features by replacing convolutional layers with capsule layers. Other common regression-based methods include deep belief network (DBN), support vector machine (SVM), extreme gradient boosting (XGBoost), random forest (RF), etc [14]. Although high-quality solutions can be obtained in a very short time, they suffer from the following limitations related to interpretation:

Submitted to the 23rd Power Systems Computation Conference (PSCC 2024).
This project is funded by the Helmholtz Association's Initiative and Networking Fund through Helmholtz AI.

- These methods are often regarded as black boxes, making it difficult to describe how load conditions are mapped to solutions.
- These methods struggle to explain the importance of power loads at each node to solutions. In other words, it is hard to explain how the power load at each node contributes to the optimal operational parameters of various power devices.

Indeed, analogous issues have been investigated beyond the RPO problem of distribution networks but within the various domains of explainable artificial intelligence (XAI) [15], [16], [17]. To the best of our knowledge, existing research primarily concentrates on image classification tasks [18], and the concept of XAI remains largely unexplored in the context of regression-based RPO of distribution networks.

In this context, this paper aims to answer the following two research questions related to the regression-based RPO of distribution networks: *How to explain the model decision of regression-based methods? How does the power load at each node contribute to the optimal operational parameters of various power devices?*

Specifically, a Shapely value theory-based explainable framework is tailored for regression-based RPO of distribution networks. Note that the proposed explainable framework is model-agnostic, and thus applicable to various regression models. The key contributions are as follows:

- Tailor a novel explainable framework based on Shapley values to tackle the black box problem, which is rarely discussed in the regression-based RPO of distribution networks.
- Propose a model-agnostic Shapley additive explanation to measure the contributions of input features (i.e., power load at each node) to solutions (i.e., operational parameters of various power devices).
- Provide two perspectives, including global perspective and instance perspective, to explain the model decision of regression-based methods by visualizations.

The rest sections are organized as follows. Section II formulates the RPO model of distribution networks. Section III shows regression-based methods in the RPO model. Section IV introduces the proposed explainable framework. Simulation and analysis are performed in section V. Finally, Section VI presents the conclusion and future work.

## II. Reactive Power Optimization Model

In the RPO model, the primary objective is to enhance power quality and minimize power loss [8]. To achieve this, a comprehensive objective function is proposed, incorporating the variations in normalized power loss and voltage deviation:

$$\max F = W_{\text{loss}}\left(P_{\text{loss}} - \hat{P}_{\text{loss}}\right)/P_{\text{loss}} + W_{\text{U}}\left(dU - d\hat{U}\right)/dU \tag{1}$$

$$dU = \frac{1}{m}\sum_{l=1}^{m}\left|\left(U_0 - U_l\right)/U_0\right| \tag{2}$$

$$P_{\text{loss}} = \sum_{i=1}^{n} G_{ij}\left[U_i^2 + U_j^2 - 2U_iU_j\cos\left(\theta_i - \theta_j\right)\right] \tag{3}$$

where $W_{\text{loss}}$ and $W_{\text{U}}$ are weights of the power loss and voltage deviation, respectively; $P_{\text{loss}}$ and $dU$ are the power loss and voltage deviation before optimization, respectively; $\widehat{P}_{\text{loss}}$ and $d\widehat{U}$ are the power loss and voltage deviation after optimization, respectively; $n$ and $m$ are the numbers of feeders and nodes, respectively; $U_0$ is the voltage base value; $U_l$ is the voltage of node $l$; $G_{ij}$ is the conductance of the feeder connecting node $i$ and node $j$; $\theta_i$ is the phase-angle of the voltage at node $i$.

In addition, the RPO model must satisfy a number of operational constraints, such as power flow constraints, voltage constraints, current constraints, and device constraints:

$$\begin{cases} P_i - P_{\text{DG},i} - U_i\sum_{j=1}^{m}U_j\left(G_{ij}\cos\theta_{ij} + B_{ij}\sin\theta_{ij}\right) = 0 \\ Q_i - Q_{\text{C},i} - Q_{\text{DG},i} - U_i\sum_{j=1}^{m}U_j\left(G_{ij}\sin\theta_{ij} - B_{ij}\cos\theta_{ij}\right) = 0 \end{cases} \tag{4}$$

$$\begin{cases} U_{\min,i} \le U_i \le U_{\max,i} & i = 1,2,\cdots m \\ I_i \le I_{\max,i} & i = 1,2,\cdots n \end{cases} \tag{5}$$

$$\begin{cases} T_{\min,i} \le T_i \le T_{\max,i} & i = 1,2,\cdots n_{\text{T}} \\ 0 \le Q_{\text{C},i} \le Q_{\text{C}\max,i} & i = 1,2,\cdots n_{\text{C}} \end{cases} \tag{6}$$

$$\begin{cases} Q_{\text{DG}\min,i} \le Q_{\text{DG},i} \le Q_{\text{DG}\max,i} \\ Q_{\text{DG}\max,i} \le \sqrt{S_{\text{DG},i}^2 - P_{\text{DG},i}^2} \end{cases} \quad i = 1,2,\cdots n_{\text{D}} \tag{7}$$

where $P_i$ and $Q_i$ are the active and reactive loads, respectively; $P_{\text{DG},i}$ and $Q_{\text{DG},i}$ are the active and reactive outputs supplied by the distribution generation $i$, respectively; $B_{ij}$ is the susceptance of the feeder connecting node $i$ and node $j$; $U_{\min}$ and $U_{\max}$ are the minimum and maximum voltage at node $i$, respectively; $I_{\max}$ is the maximum current at feeder $i$; $T_{\min}$ and $T_{\max}$ are the minimum and maximum tap position of the transformer $i$, respectively; $Q_{\text{C},i}$ is the maximum reactive output supplied by the shunt capacitor bank $i$; $n_{\text{T}}$ and $n_{\text{C}}$ are the number of transformers and shunt capacitor banks, respectively; $T_i$ and $Q_{\text{Cmax},i}$ are the tap position of the transformer $i$ and reactive output supplied by the shunt capacitor bank $i$, respectively; $Q_{\text{DGmin},i}$ and $Q_{\text{DGmax},i}$ are the minimum and maximum reactive output supplied by the distribution generation $i$, respectively; $S_{\text{DG},i}$ is the apparent power of the distribution generation $i$; and $n_{\text{D}}$ is the number of distribution generations.

To summarize, the RPO model of distribution networks is shown in Fig. 1. The operational parameters of devices to be optimized include the tap positions of transformers, reactive power outputs supplied by shunt capacitor banks, and reactive power outputs supplied by distribution generations, such as wind turbines (WT) and photovoltaic (PV) systems. Other devices (e.g. static Var compensator) have mathematical models analogous to the above and can be treated accordingly.

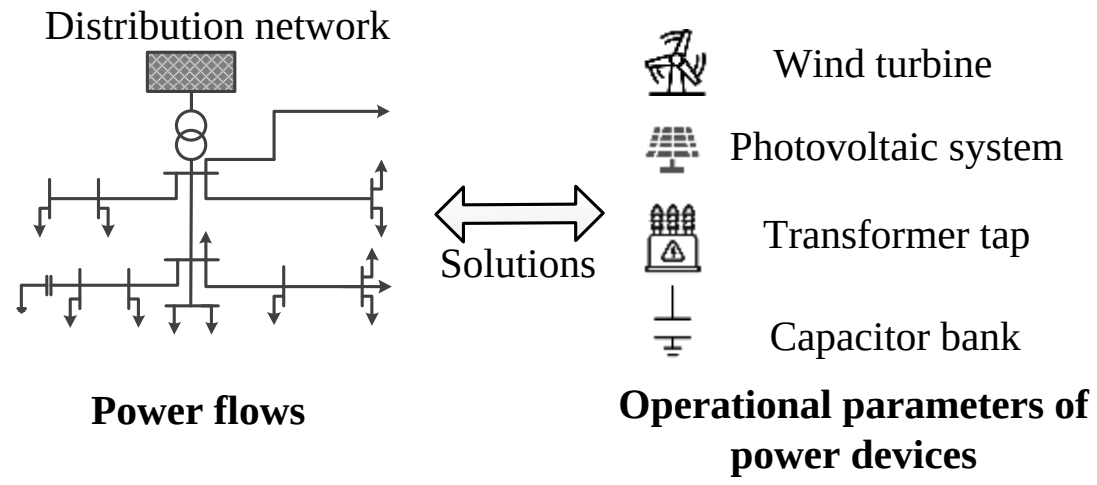

Figure 1. The RPO model of distribution networks.

## III. REGRESSION-BASED METHOD IN THE RPO MODEL

### A. *The Inputs and Outputs of the Regression-Based Method*

As shown in Fig. 2, regression-based methods can be treated as complex functions, which directly map load conditions (i.e., active and reactive power loads at nodes) to the RPO solutions (i.e., operational parameters of various power devices). Specifically, the inputs $X$ and outputs $Y$ of regression-based methods can be represented as [13]:

$$X = \left[P_2,\dots P_m;Q_2,\dots Q_m\right] \tag{8}$$

$$Y = \left[T_1,T_2,\dots T_{n_\mathrm{T}};Q_{\mathrm{C},1},Q_{\mathrm{C},2},\dots Q_{\mathrm{C},n_\mathrm{C}};Q_{\mathrm{DG},1},Q_{\mathrm{DG},2},\dots Q_{\mathrm{DG},n_\mathrm{D}}\right] \tag{9}$$

Note that node 1 is generally a slack node in the distribution network, so it does not need to be considered in the inputs. The regression-based method can be any machine learning model, such as MLP, CNN, XGBoost, RF, and SVM.

For example, MLP is able to learn the characteristics and patterns of the distribution network through a large amount of training data. They can establish a mapping between the input and output variables. During the training process, the MLP adjusts its weights and biases to minimize the error between predicted and actual outputs. When training is complete, the MLP can be used to predict and optimize the operational parameters of various power devices.

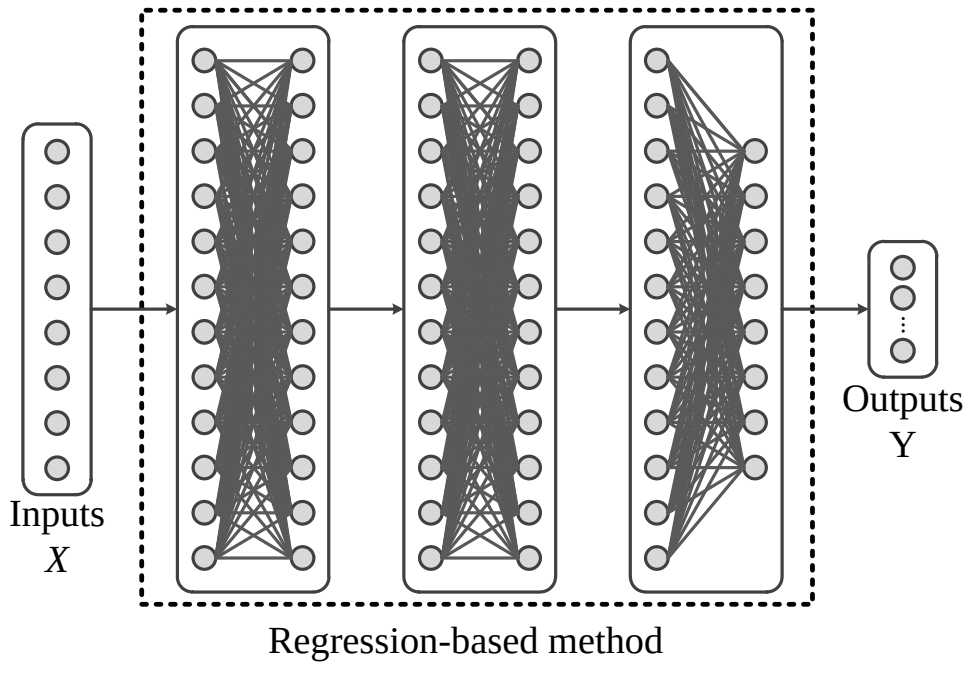

Figure 2. The regression-based method in the RPO model.

### B. *Why Does Regression-Based Method Need XAI?*

Fig. 3 briefly summarizes the reasons why XAI is needed in the RPO model of distribution networks [17].

**1) Transparent Decision**: Regression-based methods are a critical decision-making process that involves adjustments to tap positions, reactive power supplied by shunt capacitor banks, and reactive power supplied by distribution generations. Transparency is critical to the acceptance of the decision-making process. By using XAI, it can clearly demonstrate how the model makes decisions based on inputs, thereby increasing the trustworthiness and acceptance of the decisions [15].

**2) Secure Issue**: RPO plays a critical role in the security and reliability of distribution networks. The lack of explainability can hinder understanding of how the model produces optimization results, making it difficult to identify potential security risks or uncertainties. By using XAI, it can explain the model's dependence on inputs and their impact on output results, helping to identify potential problems and areas for improvement, thereby enhancing the security and reliability of distribution networks.

**3) Economic Benefit:** RPO in distribution networks involves making decisions that can affect resource allocation and have financial implications. Lack of explicability in regression-based methods can make it difficult to understand how the model arrives at certain decisions that affect resource allocation and cost effectiveness [17]. XAI allows identifying cost-saving opportunities, and makes informed decisions that maximize the economic benefits of reactive power optimization.

**4) Feature Importance:** Explanability allows distribution system operators to validate and improve machine learning-based reactive power optimization methods. By explaining the model's output results, it can identify the feature importance to the decisions, and make appropriate adjustments and improvements. As a result, the accuracy and performance of the model can be improved to better meet the needs of real-world distribution networks.

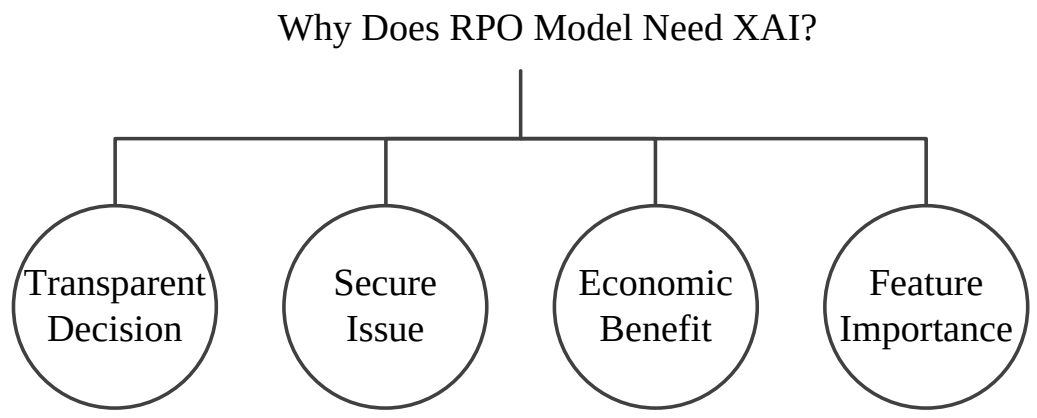

Figure 3. The reasons why XAI is needed in the RPO model.

## IV. SHAPLEY ADDITIVE EXPLANATION FRAMEWORK

### A. *The Explainable Framework*

The goal of the Shapley value-based explainable framework is to construct a post hoc explainable model, denoted $g(\cdot)$, which is capable of quantifying the contributions of features to the decision process of the original model $f(\cdot)$ by using Shapley values. This is achieved while ensuring consistency between the predicted results of the explanatory model and the original model.

Specifically, a single sample is denoted as $X=(x_1, x_2, ...,x_p)$. $p$ is the number of features. $f(X)$ is the predicted value of the original black box model, and $g(X)$ is the predicted value of the post hoc explainable model, which is usually constructed as a linear model [16]:

$$g(X) = \phi_0 + \sum_{i=1}^{p} \phi_i x_i = f(X) \tag{10}$$

where $\phi_0$ is the output of the model without considering any features; and $\phi_i$ is the Shapley value of the $i^{th}$ feature when $i$ ranges from 1 to $p$.

The contribution of a feature to the predictions, known as the Shapley value, is determined by weighting and summing over all possible combinations of feature values:

$$\phi_j = \sum_{S} \frac{(p-|S|-1)!|S|!}{p!} \left( f\left(S \cup \{j\}\right) - f\left(S\right) \right) \tag{11}$$

$$S \subseteq \{1, 2, \ldots, p\} \setminus \{j\} \tag{12}$$

where the vector $x$ represents the set of feature values associated with the sample; the subset $S$ represents a collection of features utilized within the model; $f(S)$ is the prediction of the subset $S$; and $|S|$ is the number of features in the subset $S$.

To make it easier to understand the proposed explainable framework, an example is used for visual analysis and illustration. Suppose each sample consists of 4 features denoted as $X=(x_1, x_2, x_3, x_4)$, and the original model $f(\cdot)$ is used to obtain the expectation $E(f(X))$ of the predicted values. Note that the expectation $E(f(X))$ is the same as the $\phi_0$ mentioned before. For a specific sample $Z=(z_1, z_2, z_3, z_4)$, its predicted value in the original model is $f(z)$. Then, equation (11) is used to solve the $\phi_1$, $\phi_2$, $\phi_3$, and $\phi_4$, which can explain the model.

Fig. 4 provides an explanation for the discrepancy between the individual sample predictions and the average prediction values. The blue arrows denote positive Shapley values, indicating that larger values of the corresponding features lead to increased predictions by the original model. Conversely, the pink arrows denote negative Shapley values, indicating that larger values of the corresponding features lead to decreased predictions by the original model. In particular, it is clear that features $z_1$, $z_2$, and $z_3$ are positively correlated with predictions, while $z_4$ is negatively correlated with predictions. Furthermore, $z_2$ shows the most pronounced influence on the prediction, because it possesses the largest magnitude. Ultimately, under the collective influence of these four feature variables, the predicted value of the original model for sample $z$ is moved from the mean value to $f(z)$.

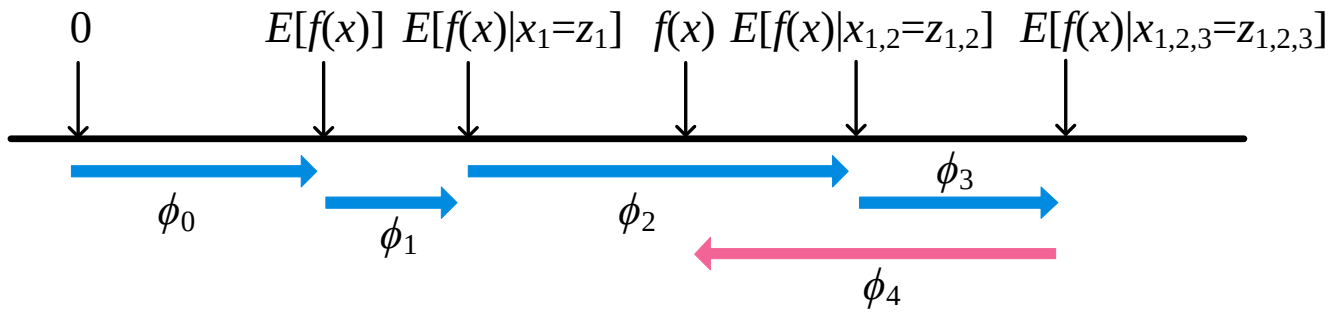

Figure 4. The discrepancy between the individual sample predictions and the average prediction values.

## B. Model-Agnostic Shapley Value Approximations

The direct computation of Shapley values requires considering all possible subsets of features, which leads to exponential growth in computational cost as the number of features increases. For cases with a large number of features or large datasets, direct computation of Shapley values can become very time-consuming or even infeasible.

To address this issue, an approximation method is presented to estimate Shapley values by employing the Gaussian kernel function to reduce the computational complexity. Specifically, the Gaussian kernel-based Shapley approximation consists of the following 3 steps [16].

**1) Define the model and features**: For any regression-based method $f(\cdot)$, its input features are marked as $X=(x_1, x_2, \ldots, x_p)$.

**2) Choose a kernel function**: A kernel function $K(\cdot)$is selected to calculate weights between samples. The Gaussian kernel is a common choice:

$$K\left(x_i, x_j\right) = \exp\left( \frac{\left\| x_i - x_j \right\|^2}{-2\sigma^2} \right) \tag{13}$$

where $\|\ \|^2$ denotes the squared Euclidean distance; and σ denotes the bandwidth parameter of the Gaussian kernel.

**3) Compute Shapley values**: A set of samples marked as $H=(X_1,X_2,\ldots,X_M)$ are selected to calculate the Shapely values. $M$ is the number of samples.

First of all, the $i^{th}$ feature is removed from the samples to obtain a new sample set marked as $H_{-i}$.

Then, the kernel weight $W_{-i}(X_j)$ between the sample $X_{-i}$ in the new sample set $S_{-i}$ and the original sample set $H$ is calculated:

$$W_{-i}\left(X_j\right) = \frac{K\left(X_{-i}, X_j\right)}{\sum_{X_j \in H} K\left(X_{-i}, X_j\right)} \tag{14}$$

Finally, the kernel weights and the outputs $f(X_{-i})$ of the new sample set $H_{-i}$ are employed to calculate the Shapley value of the $i^{th}$ feature:

$$\phi_i = \sum_{X_{-i} \in H_{-i}} W_{-i}\left(X_j\right) \cdot \left( f\left(X_{-i}\right) - f\left(X\right) \right) \tag{15}$$

In summary, after training a black-box model to obtain the solutions of the RPO problem, Eq. (15) is used to estimate Shapley values, which represent the contribution of each feature to the prediction of the regression-based method. As an approximation of Eq. (11), Eq. (15) can reduce the computational complexity. Note that the proposed Shapley value approximation is model-agnostic, and thus applicable to any regression-based method (e.g., neural networks, decision trees, SVMs, etc.) for the RPO of distribution networks.

# V. Case Study

## A. Simulation Settings

The simulations are performed on the widely used IEEE 33-node distribution network. The branch parameters are given in [19]. As shown in Fig. 5, various devices (e.g., CB, transformer, WT, and PV system) are integrated into some nodes of the distribution network.

To be specific, the voltage base value is set to 12.66 kV. The allowed voltage range for all nodes should be maintained between 0.9-1.1 p.u.. 8 CBs and 7 CBs are integrated at the 18th node and 33rd node, respectively. Each shunt CB can provide 100 kvar of reactive power. 2 WTs are positioned at the 10th and 25th nodes, and 1 PV system is positioned at the 22nd node. The capacity of both the WTs and the PV system is 500 kVA. The active power of each WT is generated by the Weibull distribution [4], while the active power of the PV system is produced by the Beta Distribution [8]. The transformer tap has a total of 17 ratios, ranging from -8×1.25% to 8×1.25%.

In the original IEEE 33-node distribution network, there is only one instance of load data, which is insufficient to adequately train and evaluate the performance of the machine learning model. To address this limitation and create a comprehensive dataset, this study introduces the assumption that the load at each node is multiplied by a noise drawn from a truncated Gaussian distribution with a mean of 1.1 and a standard deviation of 0.9, constrained within the range of 0.2 to 2 [20]. This assumption is based on the understanding that load levels often follow a Gaussian distribution, as reported in related work [20]. Subsequently, the paper generates a set of 4000 training samples (80%), along with 500 samples (10%) for validation and 500 samples (10%) for testing purposes. To construct the labels, the GA is executed 50 times, and the resulting optimal solution is assigned as the label for these respective samples.

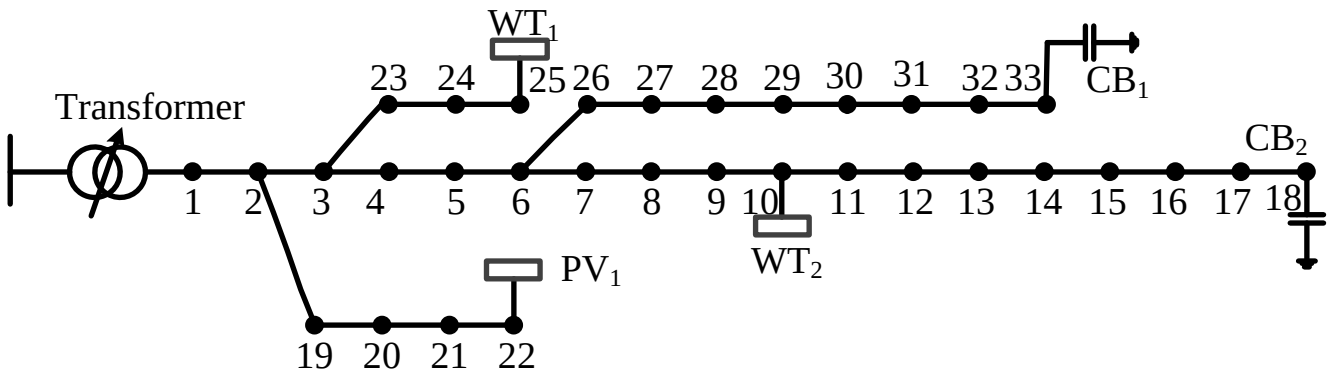

Figure 5. The topology information of IEEE 33-node distribution network.

## B. *RPO with Regression-based Method*

In this section, the widely used MLP is used as an example to conduct the RPO of distribution networks. Other black-box models can be treated in a similar way. The hyper-parameters and structures are determined through hyper-parameter optimization and exploratory experiments [21].

Specifically, a sample is randomly selected from the test set, and then the MLP maps load conditions to the operational parameters of devices. Fig. 6 shows the voltage profile of each node before and after the optimization. Note that the tap ratio before the optimization is 0.

From Fig. 6, the regression-based method achieved significant improvements in the voltage profiles of all nodes. The power loss has been reduced from 202.65 kW before optimization to 132.94 kW after optimization, while the voltage deviation has decreased from 0.052 p.u. to 0.013 p.u. after optimization. This shows the validity of the regression-based method for solving the RPO model of distribution networks.

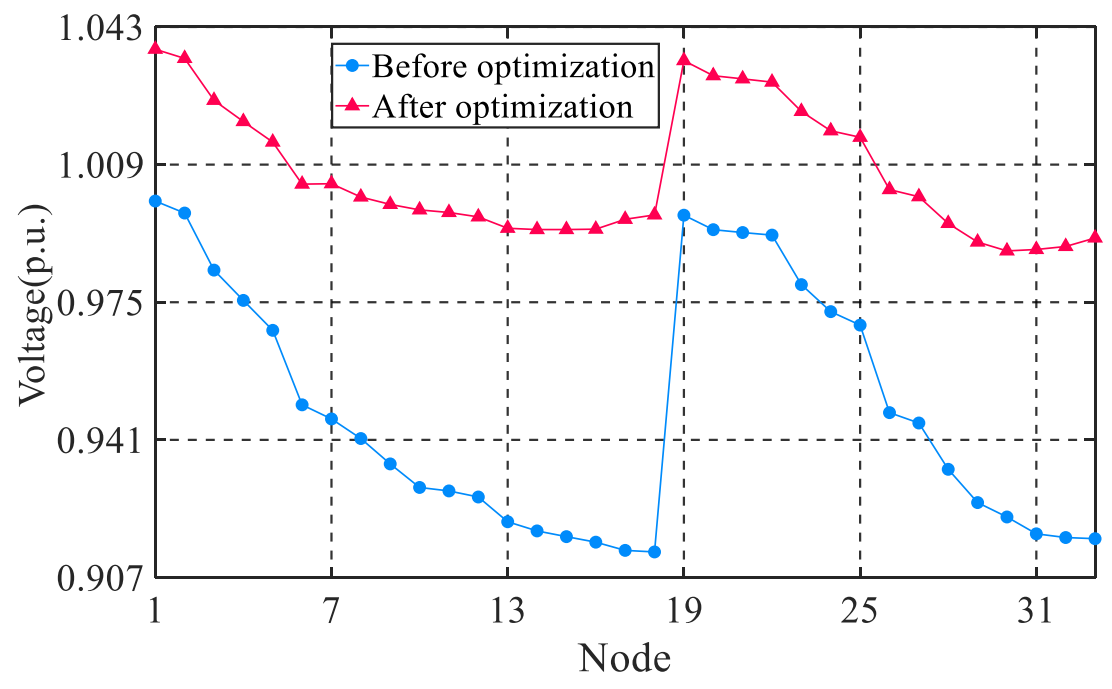

Figure 6. The voltages at nodes before and after optimization.

## C. *Explaining the Solution from a Global Perspective*

This section uses different visualizations to explain the regression-based method for the RPO of distribution networks, from a global perspective. This is important for analyzing the effect of nodal loads on the RPO.

**1) Global interpretability with bar plots**

After using MLP to obtain the solutions of the test set, Eq. (15) is used to calculate the Shapley values of the 64 features (i.e., active power loads from 2nd to node 33rd nodes and reactive power loads from 2nd to node 33rd nodes) in each sample, followed by the calculation of the average absolute Shapley value across these 64 features. Fig. 7 shows the top 20 features that have a high impact on the solution.

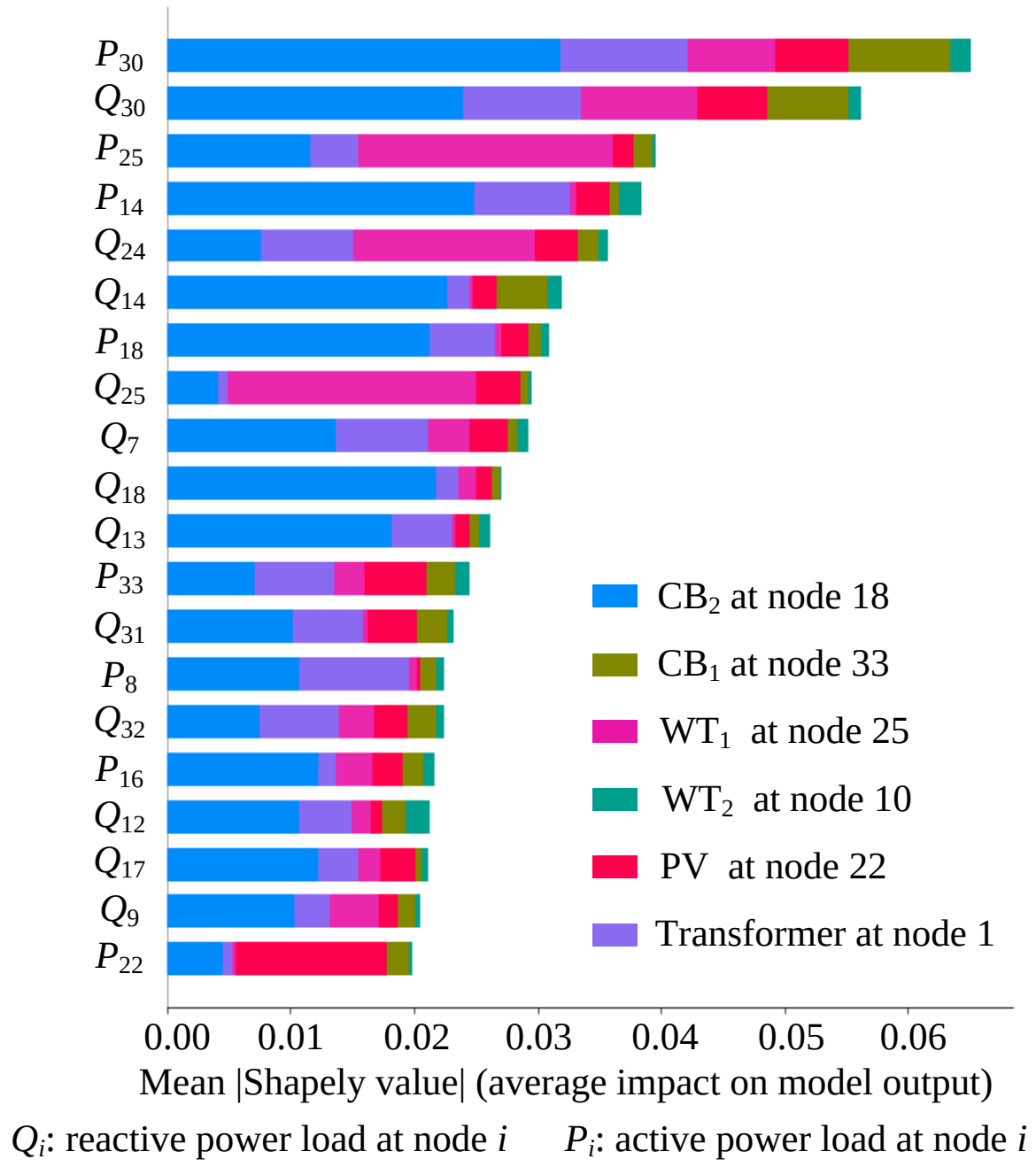

Figure 7. The top 20 features that greatly influence the solution.

The importance of each feature is represented by the lengths in the bar. When the MLP maps the features to the solution, the top five features that have the greatest impact on the solution are $P_{30}$, $Q_{30}$, $P_{25}$, $P_{14}$, and $Q_{24}$, in that order. Additionally, Fig. 7 reveals the influence of features on the device state. For instance, the red bar for feature $P_{22}$ is the longest, indicating its highest importance in determining the operational status of the PV system. This is because the reactive power supplied by the PV system is related to the load at node 22. If the load at node 22 is heavy, the PV system should supply more reactive power to keep the voltage above the lower limit. Thus, the accurate determination of the optimal parameters for the PV system based on the load at node 22 is crucial, which validates the correctness of the proposed interpretability method.

**2) Global interpretability with summary plots**

The previous bar plots employ the average absolute Shapley values to represent the feature importance, but they cannot represent the detailed Shapley value of each feature to the solution.

To solve this problem, the summary plots are adopted to provide a more comprehensive visual analysis by displaying Shapley values of each feature to solutions. As an example, Fig. 8 shows the Shapley value of each feature for the tap ratio of the transformer at node 1. The Shapley value of each feature for other devices can be explained in a similar way.

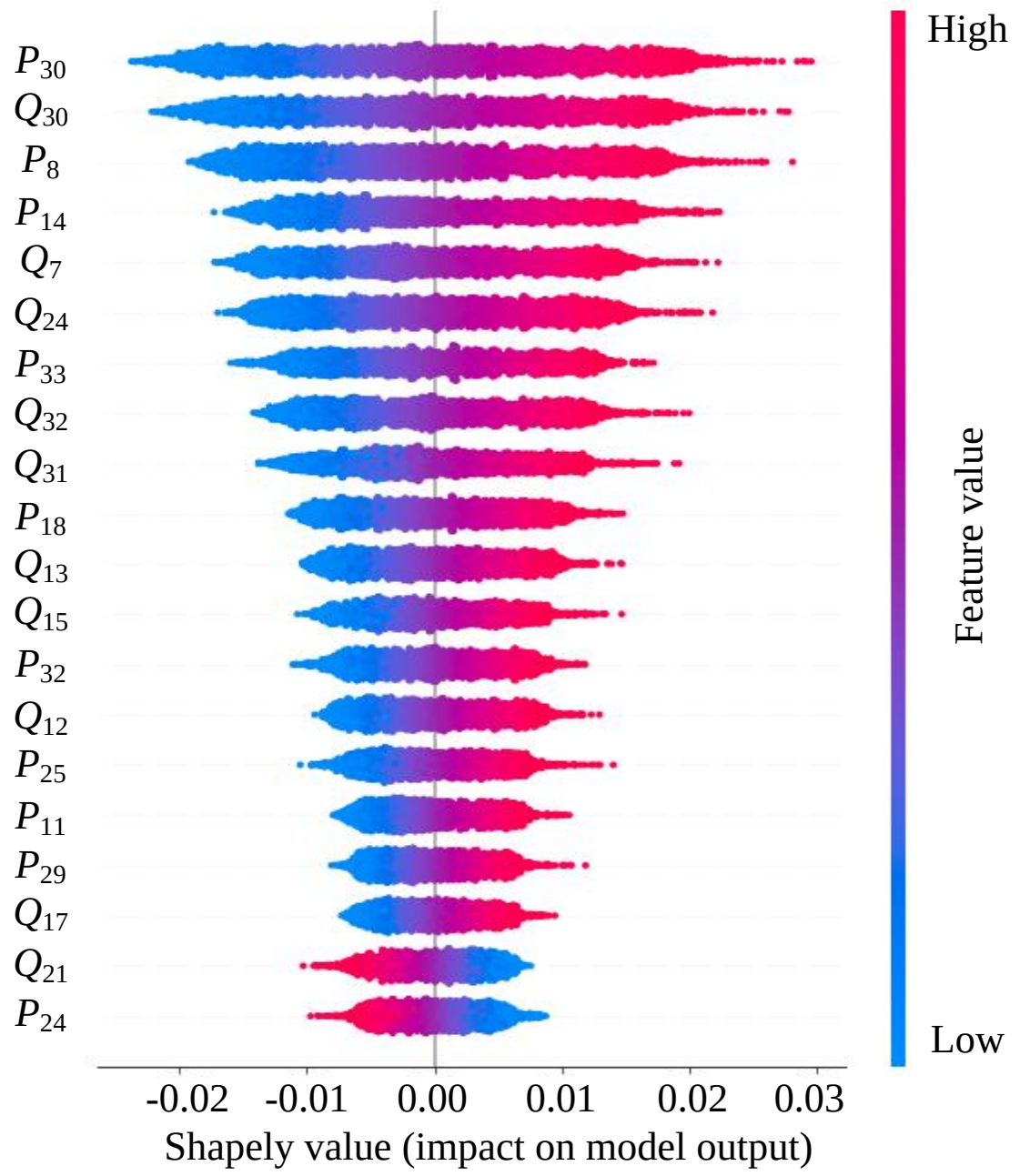

$Q_i$: reactive power load at node $i$ $P_i$: active power load at node $i$

Figure 8. The summary plots of the Shapley values for all samples.

The summary plot represents the descending order of the feature importance as determined by the Shapley values. The use of horizontal violin plot visualization effectively illustrates the positive or negative impact of each feature across different data points. In addition, the color combinations used in the plot provide a clear indication of the relative impact of the feature values from high to low.

The color displayed in the horizontal violin plots for each feature indicates whether there is a positive or negative association with the solution (i.e., operational parameters of devices). The red color represents a positive association, while the blue color represents a negative association.

It is observed that a significant portion of the features in the SHAP summary plots had approximately equal Shapley values of red color and blue color. This indicates that these features have a roughly balanced positive and negative impact on the solution. This is because there is a comparable number of samples for both heavy-load and light-load scenarios in the dataset. Light-load conditions result in negative Shapley values, indicating that the transformer tends to set a small tap ratio to prevent the voltage from crossing the upper limit. On the other hand, heavy-load conditions lead to positive Shapley values, implying a need for the transformer to set a large tap ratio to avoid the voltage falling below the lower limit.

**3) Global interpretability with feature dependence plot**

The dependence plot examines the relationship between a specific feature and another feature by assessing their partial dependence. To illustrate this, Fig. 9 displays the dependence plot between the reactive power load and the active power load at node 18. The visualization explains that there is a positive correlation between the active and reactive loads at the nodes, aligning with the load patterns typically observed in distribution networks.

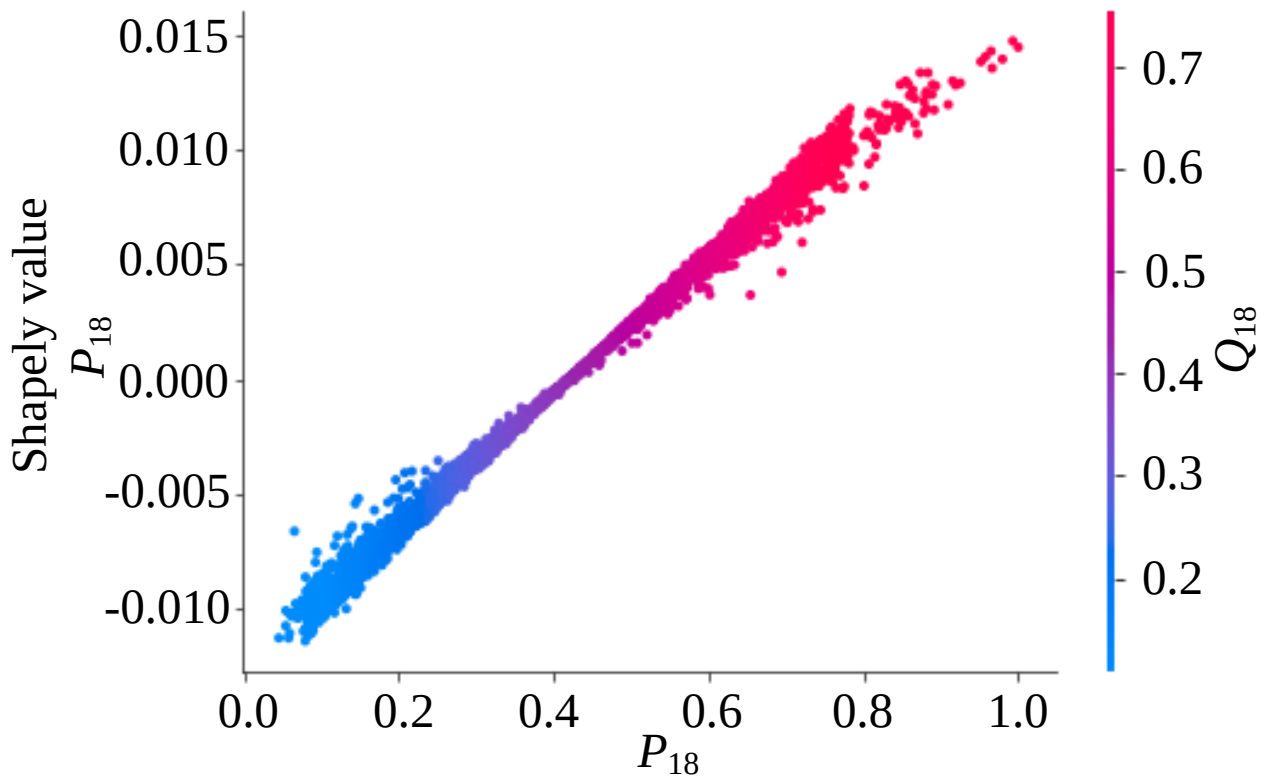

$Q_i$: reactive power load at node $i$ $P_i$: active power load at node $i$

Figure 9. The interaction between the reactive power load and the active power load at node 18.

Normally, when the active power load at a node is heavy, it indicates a significant consumption of electrical energy by the devices connected to that node. Consequently, there is a higher likelihood of a substantial reactive power load as well. This is attributed to the need for compensating reactive power to maintain a stable power factor and voltage within acceptable limits, thereby balancing the demand for active power.

Overall, dependence plots provide a way to explain interactions among influential features in the distribution networks.

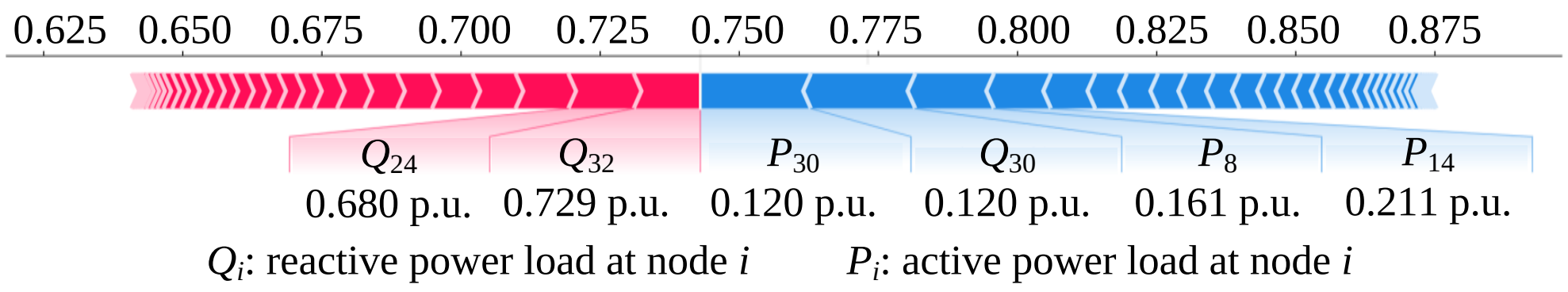

Figure 10. The force plot of a randomly selected instance.

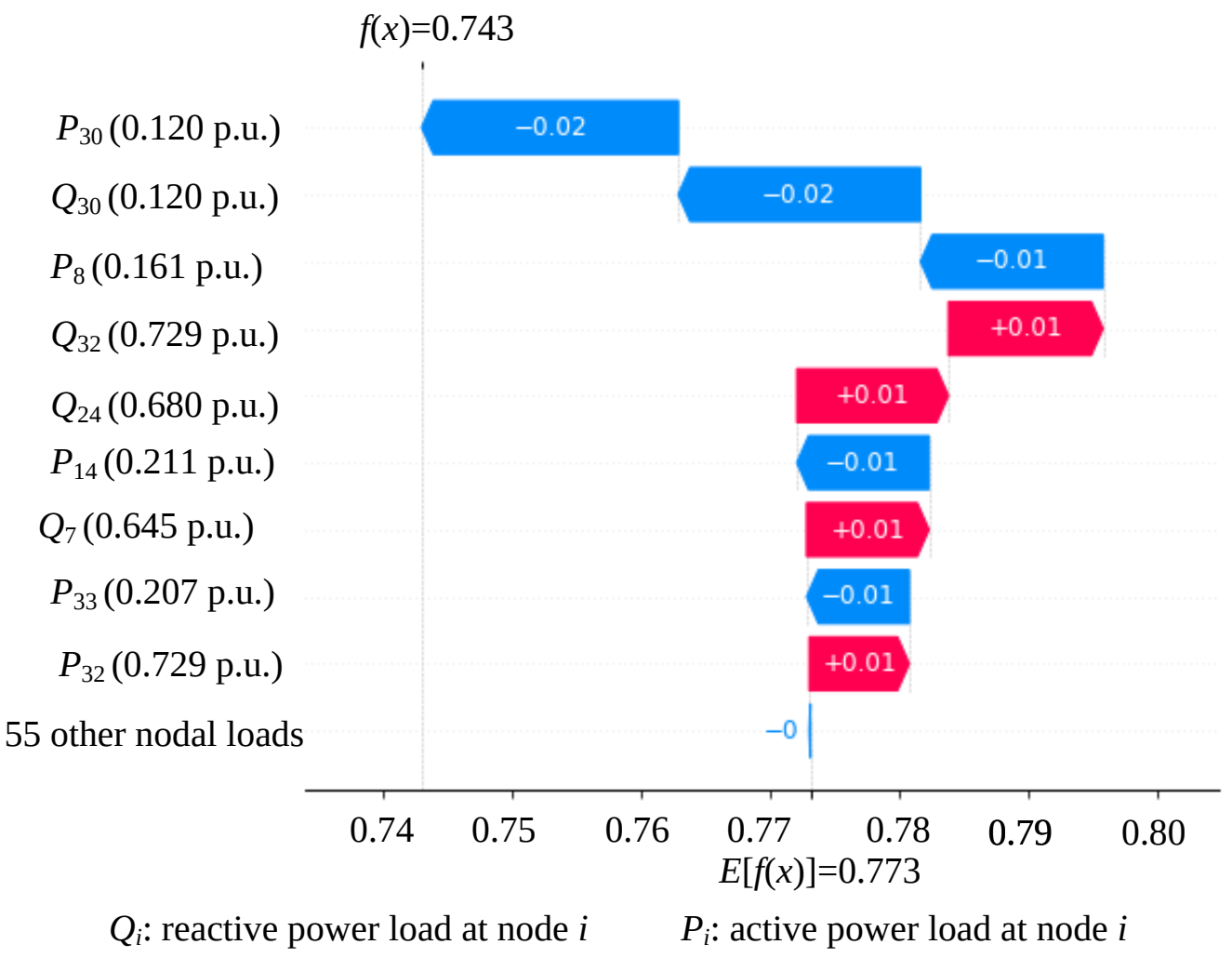

Figure 11. The waterfall plot of a randomly selected instance.

### D. *Explaining the Solution from an Instance Perspective*

This section uses different visualizations to explain the regression-based method for the RPO of distribution networks, from an instance perspective. Due to the page limit, the Shapley value of each feature for the tap ratio of the transformer at node 1 is considered as an example. The Shapley value of each feature for other devices can be explained in a similar way.

**1) Instance interpretability with force plots**

To explain the feature importance in a specific instance, a sample is randomly selected, and then force plots are used to show the Shapley values of the features for the tap ratio of the transformer at node 1, as shown in Fig. 10.

For this specific instance, the $Q_{24}$, $Q_{32}$, $P_{30}$, $Q_{30}$, $P_8$, and $P_{14}$ greatly influence the solution. Besides, the force plot shows each Shapley value as an arrow, indicating its effect on the tap ratio, either pushing it up (positive value) or down (negative value). From Fig. 10, $P_{30}$, $Q_{30}$, $P_8$, and $P_{14}$ decrease the tap ratio, while $Q_{24}$ and $Q_{32}$ increase the tap ratio. This is because $P_{30}$, $Q_{30}$, $P_8$, and $P_{14}$ are light power loads, and the transformer ensures that the voltage at these nodes does not exceed the upper limit. In contrast, $Q_{24}$ and $Q_{32}$ are heavy loads or medium loads, and thus the transformer should provide a large tap ratio to maintain the voltage.

**2) Instance interpretability with waterfall plot**

The waterfall plot serves as an alternative interpretability that offers advantages over bar plots. Unlike bar plots, the waterfall plot provides detailed insight into how each feature value in a particular instance contributes positively or negatively to the solution. As an example, a sample is randomly selected, and then the waterfall plot presents how each nodal load contributes positively or negatively to the tap ratio, as shown in Fig. 11.

For this specific instance, the negative Shapley values of $P_{30}$, $Q_{30}$, $P_9$, $P_{14}$, and $P_{33}$ indicate that they are negatively influencing the model to give a smaller tap ratio. This explanation is reasonable and consistent with practical expectations, as it is logical to reduce the tap ratio to prevent the voltage from exceeding the upper limit under light load conditions ($P_{30}$, $Q_{30}$, $P_9$, $P_{14}$, and $P_{33}$ are light load conditions).

On the contrary, the positive Shapley values of $Q_{32}$, $Q_{24}$, $Q_7$, and $P_{32}$ indicate that they are positively influencing the model to give a larger tap ratio for this specific instance. Similarly, it is rational to adjust the tap ratio to a higher value in order to prevent the voltage from falling below the lower limit under heavy load conditions (specifically, $Q_{32}$, $Q_{24}$, $Q_7$, and $P_{32}$).

### E. *Can We Trust the Explanation?*

In this section, simulations are performed to analyze whether the explanation provided by the proposed method can be trusted. Due to the page limit, the Shapley value of each feature for the tap ratio of the transformer at node 1 is still considered as an example. The Shapley value of each feature for other devices can be explained in a similar way.

Firstly, for each sample in the test set, the Shapley values of individual features (i.e., nodal load) for the tap ratio are computed. Note that a positive Shapley value indicates that the feature tends to increase the tap ratio (they are positively correlated), while a negative Shapley value indicates that the feature tends to decrease the tap ratio. Next, a threshold is set to mark features (i.e., nodal loads) as light or heavy. If the explanation is reliable, the Shapley values of features marked as light should be negative, while the Shapley values of features marked as heavy should be positive. Finally, Fig. 12(a) presents the markers of each sample in the test set, while Fig.12 (b) displays the positive and negative Shapley values of features in each sample for the tap ratio in the test set.

Fig. 12 (a) resembles very well to Fig. 12(b), with 97.44% of the elements being identical. In other words, most of the features labeled as light have negative Shapley values, while the features labeled as heavy have positive Shapley values. This suggests that the explanation provided by the proposed method is trustworthy.

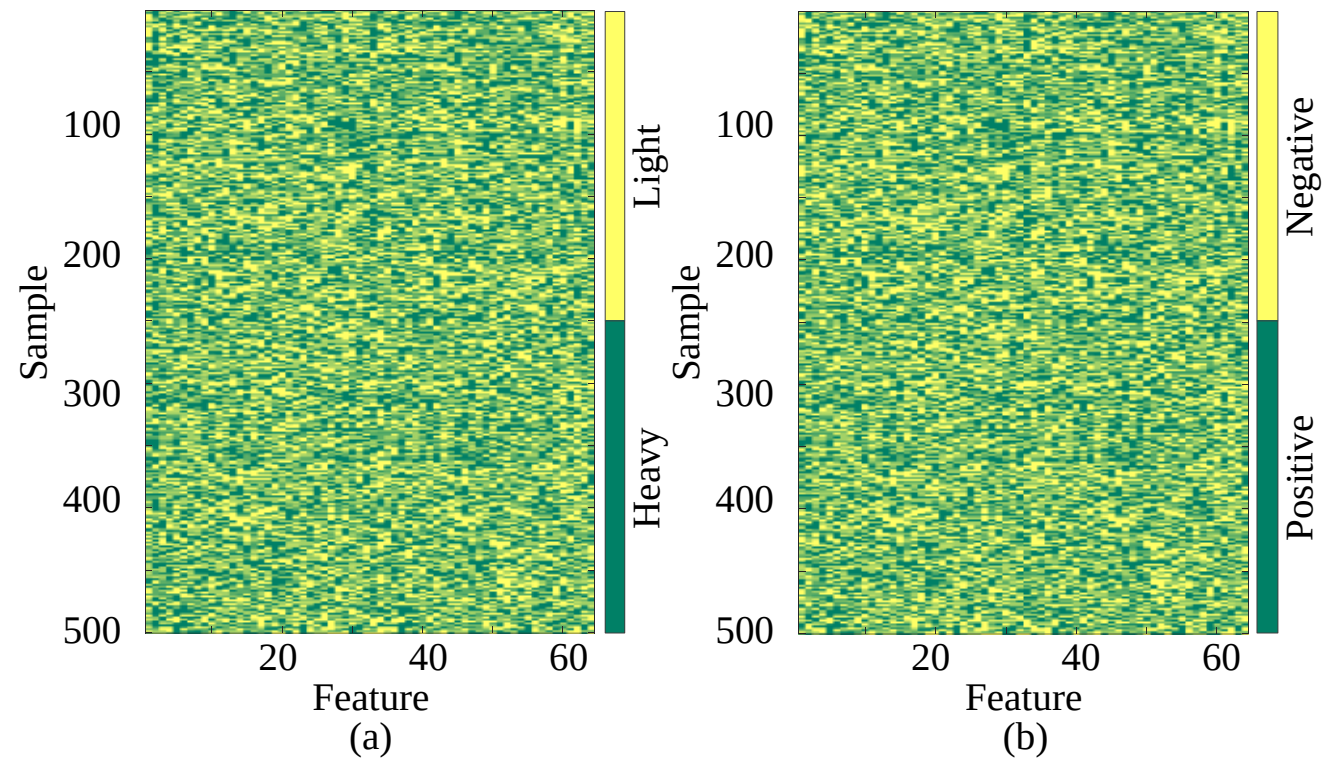

Figure 12. The markers and Shapely values of features.(a) Markers of each sample. (b) Shapley values of features in each sample for the tap ratio.

## VI. Conclusion

The regression-based methods for solving the RPO model are often regarded as black boxes, which is hard to explain the relationship between load conditions and solutions. To address this problem, a novel explainable framework based on Shapley values is tailored for regression-based RPO of distribution networks. Simulation and analysis on the IEEE 33-node distribution network lead to the following conclusions:

1) The regression-based method achieves good performance. Specifically, power loss and voltage deviation are significantly reduced by directly mapping load conditions into solutions (i.e., operational parameters of the device).

2) XAI makes black box models transparent. In particular, from a global perspective, the proposed method not only provides the importance ranking of nodal loads for solutions, but also analyzes the relationship between nodal loads and solutions. In addition, the interactions between node loads can be visualized by using the feature dependency plot. From an instance perspective, the proposed method allows the analysis of the impact of each nodal load on the operational parameters of devices given a specific sample. Generally, this transparency not only builds the trust and acceptance of the solutions from regression-based methods, but also gives insights into which parts of nodal loads drive instability.

3) The comparison between the load patterns and Shapely values suggests that the interpretation provided by the proposed method is trustworthy.

The proposed method is only applicable to regression-based methods (e.g., neural networks). XAI for similarity-based RPO can be explored in extended work.

## References

[1] H. Singh, Y. Sawle, S. Dixit, H. Malik, and F. Marquez, "Optimization of reactive power using dragonfly algorithm in DG integrated distribution system," *Electr. Power Syst. Res.*, vol. 220, pp. 1-10, Jul. 2023.

[2] T. Ibrahim, T. T. D. Rubira, A. D. Rosso, M. Patel, S. Guggilam and A. A. Mohamed, "Alternating Optimization Approach for Voltage-Secure Multi-Period Optimal Reactive Power Dispatch," *IEEE Trans. Power Syst.*, vol. 37, pp. 3805-3816, Sept. 2022.

[3] Y. Levron, Y.Beck, L. Katzir, and J. Guerrero, "Real-time reactive power distribution in microgrids by dynamic programing," *IET Gener. Transm. Distrib.*, vol. 11, pp. 530-539, Jan. 2017.

[4] Y. Tang, H. He, Z. Ni, J. Wen, and X. Sui, "Reactive power control of grid-connected wind farm based on adaptive dynamic programming," *Neurocomputing.*, vol. 125, pp. 125-133, Feb. 2014.

[5] D. Hu, Z. Ye, Y. Gao, Z. Ye, Y. Peng and N. Yu, "Multi-Agent Deep Reinforcement Learning for Voltage Control With Coordinated Active and Reactive Power Optimization," *IEEE Trans. Smart Grid.*, vol. 13, pp. 4873-4886, Nov. 2022.

[6] R. Hu, W. Wang, X. Wu, Z. Chen, L. Jing, W. Ma, and G. Zeng, " Coordinated active and reactive power control for distribution networks with high penetrations of photovoltaic systems," *Sol Energy.*, vol. 231, pp. 809-827, Jan. 2022.

[7] H. Kuang, F. Su, Y. Chang, K. Wang, and Z. He, "Reactive power optimization for distribution network system with wind power based on improved multi-objective particle swarm optimization algorithm," *Electr. Power Syst. Res.*, vol. 213, pp. 1-6, Dec. 2022.

[8] Q. Zhao, W. Liao, S. Wang and J. R. Pillai, "Robust Voltage Control Considering Uncertainties of Renewable Energies and Loads via Improved Generative Adversarial Network," *J. Mod. Power Syst. Clean Energy.*, vol. 8, pp. 1104-1114, Nov. 2020.

[9] J. Wu, C. Shi, M. Shao, R. An, X. Zhu, X. Huang, and R. Cai, "Reactive Power Optimization of a Distribution System Based on Scene Matching and Deep Belief Network," *Energies*, vol. 12, pp. 1-24, Aug. 2019.

[10] W. Liao, S. Wang, Q. Liu, and X. Shu, "Reactive Power Optimization of Distribution Network Based on Case-Based Reasoning," *2018 IEEE Power & Energy Society General Meeting (PESGM)*, Portland, USA, pp. 1-5, Aug. 2018.

[11] M. Shahriyari, A. Safari, A. Quteishat, and H. Afsharirad, "A short-term voltage stability online assessment based on multi-layer perceptron learning," *Electr. Power Syst. Res.*, vol. 223, pp. 1-14, Oct. 2023.

[12] H. Samet, S. Ketabipour, S. Afrasiabi, M. Afrasiabi, and M. Mohammadi, "Prediction of wind farm reactive power fast variations by adaptive one-dimensional convolutional neural network," *Comput. Electr. Eng.*, vol. 96, pp. 1-16, Dec. 2021.

[13] W. Liao, J. Chen, Q. Liu, R. Zhu, L. Song and Z. Yang, "Data-driven Reactive Power Optimization for Distribution Networks Using Capsule Networks," *J. Mod. Power Syst. Clean Energy.*, vol. 10, pp. 1274-1287, Sept. 2022.

[14] Z. Zhang, S. Tang, and Y. Sun, "Multi-time scale adaptive reactive power and voltage control of distribution network based on random forest algorithm," *Proceedings of the 3rd Asia-Pacific Conference on Image Processing, Electronics and Computers*, Dalian, China, pp. 1060–1065, Apr. 2022.

[15] A. Mehinovic, S. Grebovic, A. Fejzic, N. Oprasic, S. Konjicija, A. Aksamovic, "Application of artificial intelligence methods for determination of transients in the power system," *Electr. Power Syst.* Res., vol. 223, pp. 1-7, Oct. 2023.

[16] G. Mitrentsis, H. Lens, " An interpretable probabilistic model for short-term solar power forecasting using natural gradient boosting," *Appl. Energy*, vol. 309, pp. 1-20, Mar. 2022.

[17] R. Machlev, L. Heistrene, M. Perl, K. Levy, J. Belikov, S. Mannor, and Y. Levron, "Explainable Artificial Intelligence (XAI) techniques for energy and power systems: Review, challenges and opportunities," *Energy AI.*, vol. 9, pp. 1-13, Aug. 2022.

[18] N. I. Arnold, P. Angelov and P. M. Atkinson, "An Improved Explainable Point Cloud Classifier (XPCC)," *IEEE Trans. Artif.*, vol. 4, pp. 71-80, Feb. 2023.

[19] M. Baran and F. Wu, "Network reconfiguration in distribution systems for loss reduction and load balancing," *IEEE Trans. Power Deliv.*, vol. 4, 1401-1407, Apr. 1989.

[20] M. Doostizadeh, and H. Ghasemi, "Day-ahead scheduling of an active distribution network considering energy and reserve markets," *Int. Trans. Electr. Energ. Syst.*, vol. 23, pp. 930-945, Oct. 2013.

[21] L. Yang and A. Shami, "On hyperparameter optimization of machine learning algorithms: Theory and practice," *Neurocomputing.*, vol. 415, pp. 295-316, Nov. 2020.